\documentstyle[11pt,fleqn]{article}
\begin{document}

\title {\large A COLORED PARTICLE ACCELERATION BY FLUCTUATIONS IN QGP
\thanks {This work is supported by the National Natural Science Fund of China.}}
\author{{Z{\small HENG} X{\small IAOPING}{\hskip .5cm}
L{\small I} J{\small IARONG}}\\
{\small The Institute of Particle Physics of Huazhong Normal University}\\
{\small Wuhan,430070,P.R.China}}
\maketitle

\vskip-8truecm
\hskip12truecm HZPP-97-04

\hskip12truecm July 1, 1997
\vskip6truecm

\begin{center}
\begin{minipage}{120mm}
\begin{center}{\bf ABSTRACT}\end{center}
{ \ \ \ We discuss the energy variation of a parton passing through a
quark-gluon plasma(QGP) taking into account nonlinear polarization effect.
We find the parton can be accelerated by fluctuations in QGP, which
gives us a new physical insight about the response of QGP to such external
the current.
}
\end{minipage}
\end{center}

\vfill
{\large\bf
\begin{center} INSTITUTE OF PARTICLE PHYSICS
\vskip0.2truecm

HUAZHONG NORMAL UNIVERSITY
\vskip0.2truecm

WUHAN CHINA \end{center}
}
\newpage

\indent
There exist hard and soft processes in relativistic heavy ion collisions.
The former cause to produce the fast particles such as $J/\psi$, the later
can lead to thermal dense matter such as quark-gluon plasma (QGP). When
$J/\psi$ passes through  QGP, on one hand, it is well known fact that a
$J/\psi$ jet is suppressed because the deconfinemental transition can dissolve
the $J/\psi$ particle into partons, on the other hand, the fast partons will
interact with the QGP and the energy of the partons should be changed.
Since Satz et al
pointed out that $J/\psi$-suppression is a feasible sign to probe a QGP in
1986\cite{r1}, a number of experiments and theoretical analyses have been made. 
Up to now, however, one has not 
determined whether  QGP results from the heavy ion collisions.
Some works\cite{r2,r3,r4,r5} have been studying the energy loss of a fast parton passing
through  QGP in recent years. One expected the $J/\psi$-suppression
together energy variations of partons can be used to define the real signs
about the production of QGP. 

\indent
Now we set about our analyses and calculations. Without considering
collisions of the parton with individual particles in  QGP, the classical
expression for the parton energy variation pur unit time reads
\begin{equation}
(\frac {{\rm d}W}{{\rm d}t})=\int {\rm d}^3x{\bf j}^a_{\rm ext}(x)\cdot {\bf E}^a(x),
\end{equation}
where ${\bf E}^a, a=1,\cdots 8$ is the chromoelectric field induced in
the plasma by the external current ${\bf j}^a_{\rm ext}$. Suppose a parton with
charge $Q^a$ is moving in the plasma with a fixed velocity ${\bf v}$.
The external current density created by the parton charge is
\begin{equation}
{\bf j}^a_{\rm ext}(x)=Q^a{\rm v}\delta({\bf x}-{\bf v}t),
\end{equation}
$Q^aQ_a=g^2C_Q$ where $C_Q={4\over 3}(3)$ for a quark(gluon); $g$ is
 the QCD coupling constant, $\alpha_s=g^2/4\pi$.
In a linear response theory, one has completed the calculus of Eq(1) and
found that the work done by the induced fields on the external current created
by moving parton is always negative, namely, a parton loses its energy.

\indent
As is well known, the nonlinear effects arises from fluctuations in a plasma
are important, and
they influence undoubtedly on the calculations of Eq(1). It has been verified
that the linear effects are Abelian-like and pure non-Abelian effect dominates
only in the nonlinear effects\cite{r7,r8}. Therefore, this paper emphasizes
the solutions for Eq(1) in the framework of a nonlinear response theory.

\indent
According to the kinetic theory for QGP, the fluctuation currents $j_T$ can be
defined by the fluctuations of particle distributions and expanded in terms of
fluctuate field $E^T$. We had ever proved that the nonlinear effects are 
observable until three-order current\cite{r7,r6}. Thus, the field equation
in QGP reads
\begin{equation}
-\omega\epsilon^l(\omega,k)[E(\omega,k)+E^T(\omega,k)]
	  ={\rm i}{{\bf k}\over k}\cdot({\bf j}^{(3)}_T(\omega,k)+{\bf j}_{\rm ext}(\omega, k)).
\end{equation}
where
\begin{equation}
\epsilon^l(\omega, k)=1+{m_D^2\over k^2}
		  \left [
		   1+{\omega\over 2k}{\rm ln}
		    \vline
		      {\omega-k \over \omega +k}
		    \vline
		   +{{\rm i}\pi\omega\over 2k}\theta(k-|\omega|)
		  \right ]
\end{equation}
is the longitudinal dielectric coefficient of QGP in a linear response
approximation. Multiplying Eq(3) by $E(\omega', k')+E^T(\omega',k')$, 
taking the average 
over phases and integrating over $\omega'$ and $k'$, we have
\begin{equation}
-\omega [\epsilon^l(\omega, k)+\epsilon^n(\omega, k)][E^2(\omega,k)+I(\omega,k)]
={\rm i}{{\bf k}\over k}\cdot {\bf j}_{\rm ext}(\omega,k)E(\omega,k),
\end{equation}
where $\epsilon^n$ is called nonlinear permeability\cite{r6}; $I=<E^TE^T>$
is fluctuation correlation intensity.
Computing $\epsilon^n$ to the  leading
order in $g$, we find\cite{r7}
\begin{equation}
\epsilon^n(\omega,k)=-{Ng^2m_D^2\over 4\omega^2}\int d^4k_1I({\bf k_1})
	      {\pi\over\omega_1^2}\delta(\omega_1-\omega_1({\bf k_1}))
	      (\omega\Gamma_{k,k_1,-k_1,k}+\omega_1\Gamma_{k,k_1,k,-k_1})
\end{equation}
$$\Gamma_{k,k_1,k_2,k_3}=\int {d\Omega\over 4\pi}{{\bf v}_p{\bf k}\over |{\bf k}|}{{\bf v}_p{\bf k}_1\over |{\bf k}_1|}
      {{\bf v}_p{\bf k}_2\over |{\bf k}_2|}{{\bf v}_p{\bf k}_3\over |{\bf k}_3|}{1\over v_pk+i0^+}{1\over
      v_p(k_2+k_3)+i0^+}{1\over v_pk_3+i0^+}.$$
where ${\bf v}_p$ is thermal particle velociety. 
We know the pure non-Abelian contributions to the nonlinear permeability
dominate in QGP in accordance with Ref[7]. Since the fluctuate field $E^T$ is
weak, we can consider $I\ll E^2$. Thus, Eq(5) is able to simplified as
\begin{equation}
-\omega[\epsilon^l(\omega,k)+\epsilon^n(\omega,k)]E(\omega,k)
={\rm i}{{\bf k}\over k}\cdot{\bf j}_{\rm ext}.
\end{equation}

Therefore, we are able to obtain the field created by a current from Eq(7)
under taking nonlinear polarization into account.
Substituting the chromoelectric field found from Eq(7) and the Fourier 
transformed current Eq(2) into Eq(1), one finds
\begin{equation}
\left (\frac{{\rm d}W}{{\rm d}t}\right )
               ={C_Q\alpha_s\over 2\pi^2}
	       \int {\rm d}^3k{\rm d}\omega\omega
		 \left [
		  {1\over k^2}{\rm Im}{1\over \epsilon^l+\epsilon^n}
		 \right ]
	       \delta(\omega-{\bf v\cdot k}),
\end{equation}
It is convenient to change the energy variation pur unit time into the energy
variation pur unit length:
\begin{equation}
\left (\frac{{\rm d}W}{{\rm d}x}\right )
	     ={C_Q\alpha_s\over 2\pi^2v}
	       \int {\rm d}^3k{\rm d}\omega\omega
		 \left [
		  {1\over k^2}{\rm Im}{1\over \epsilon^l+\epsilon^n}
		 \right ]
	       \delta(\omega-{\bf v\cdot k}),
\end{equation}
where $v= |{\bf v}|$. In weak turbulent theory, Re$\epsilon^l\gg {\rm Re}
\epsilon^n$, we write respectively the variations due to linear effects 
and nonlinear effects following
\begin{equation}
\left (\frac{{\rm d}W}{{\rm d}x}\right )^l
	     =-{C_Q\alpha_s\over 2\pi^2v}
	       \int {\rm d}^3k{\rm d}\omega\omega
		 \left [
		  {1\over k^2}{{\rm Im}\epsilon^l\over (\epsilon^l)^2}
		 \right ]
	       \delta(\omega-{\bf v\cdot k}),
\end{equation}
and
\begin{equation}
\left (\frac{{\rm d}W}{{\rm d}x}\right )^n
	     =-{C_Q\alpha_s\over 2\pi^2v}
	       \int {\rm d}^3k{\rm d}\omega\omega
		 \left [
		  {1\over k^2}{{\rm Im}\epsilon^n\over (\epsilon^l)^2}
		 \right ]
	       \delta(\omega-{\bf v\cdot k}).
\end{equation}
Eq(10) represents the energy variation of the parton due to the linear 
polarization effects of QGP and have been discussed in\cite{r4,r5}. 
Since the right side 
of Eq(10) is always negative, ones spontaneously reach a conclusion that 
the linear polarization effects lead to the energy loss of the parton\cite{r2,r4,r5}.
Unlike the previous studies, we obtain another energy variation (Eq(11))  
besides the known result (Eq(10)) in this letter. The new energy variation
of the parton is due to nonlinear polarization arising from the interactions 
of fluctuations in QGP. 

\indent
To clarify deeply the point into the essence of Eq(11), we must solve Eq(11). 
Substituting Eq(6)
into Eq(11), we arrive at
\begin{eqnarray}
\left ({{\rm d}W\over {\rm d}x}\right )^n 
   &=&
     \frac{NC_Q\alpha_s^2m_D^2}{8\pi^3v}
       \int\frac{{\rm d}^3k}{k^2}\frac{{\rm d}\omega}{\omega}
           {1\over (\epsilon^l)^2}\delta(\omega-{\bf v\cdot k})
        \int{\rm d}^3k_1I(k_1){1\over\omega^2_1}\nonumber\\
         &\times&\int{{\rm d}\Omega\over 4\pi}{{\bf v}_p\cdot{\bf k}\over k^2}
                                      {{\bf v}_p\cdot{\bf k_1}\over k_1^2}
                                {\cal P}(\omega_1{\cal P}_1-\omega{\cal P})
                          \delta\left (
                                 \omega-\omega_1-{\bf v}_p\cdot ({\bf k-k}_1)
                                \right ),
\end{eqnarray}
where ${\cal P}={\cal P}{1\over \omega-{\bf v}_p\cdot{\bf k}}, 
{\cal P}_1={\cal P}{1\over \omega_1-{\bf v}_p\cdot{\bf k}_1};$
${\cal P}$ denotes taking the principal value of a function.
To obtain some analytic results, it is necessary to divide the integral over 
the transferred momentum $k$
into two distinct regions: short wavelength($k> k_0$) and long wavelength
($k< k_0$), where $k_0$ is a momentum of the order of the Debye screening
mass $m_D$. Since the factor $(\omega_1{\cal P}_1-\omega{\cal P})$ tend to 
zero when the transferred momentum $k$ is comparable to the momentum of 
fluctuation mode $k_1$, the energy variation is important only when $k$ and
$k_1$ are in different wavelength region. Thus, the energy variation in Eq(12)
is expressed as a summation of two parts:
\begin{equation}
\left ({{\rm d}W\over {\rm d}x}\right )^n =\left ({{\rm d}W\over {\rm d}x}\right )^n_{k>k_0}
+\left ({{\rm d}W\over {\rm d}x}\right )^n_{k<k_0}, 
\end{equation}
where
\begin{eqnarray}
\left ({{\rm d}W\over {\rm d}x}\right )^n_{k>k_0}
   &=&
     \frac{NC_Q\alpha_s^2m_D^2}{8\pi^3v}
       \int_{k_0}^{k_{\rm max}}\frac{{\rm d}^3k}{k^2}\frac{{\rm d}\omega}{\omega}
           \delta(\omega-{\bf v\cdot k})
   \int_0^{k_0}{\rm d}^3k_1I(k_1){1\over\omega^2_1}\nonumber\\
           &\times&\int{{\rm d}\Omega\over 4\pi}\cos ^2\alpha\cos ^2\beta
                                {\cal P}(\omega_1{\cal P}_1-\omega{\cal P})
                          \delta (\omega-\omega_1-k\cos\alpha ),
\end{eqnarray}
\begin{eqnarray}
\left ({{\rm d}W\over {\rm d}x}\right )^n_{k<k_0}
 &=&
     \frac{NC_Q\alpha_s^2m_D^2}{8\pi^3v}
       \int_0^{k_0}\frac{{\rm d}^3k}{k^2}\frac{{\rm d}\omega}{\omega}
           {1\over (1+{m_D\over k^2})^2}\delta(\omega-{\bf v\cdot k})
 \int_{k_0}^{k_{\rm max}}{\rm d}^3k_1I(k_1){1\over\omega^2_1}\nonumber\\
         &\times&\int{{\rm d}\Omega\over 4\pi}\cos ^2\alpha\cos ^2\beta
                                {\cal P}(\omega_1{\cal P}_1-\omega{\cal P})
                                \delta (\omega-\omega_1+k_1\cos\beta).
\end{eqnarray}
Here, we consider massless plasma particles; It is a good approximation to
set $\epsilon^l=1$ in the short wavelength region and 
$\epsilon^l=1+{m_D^2\over k^2}$ in the long wave region(see Eq(4)) since 
the momentum 
provided by the external current is always space like\cite{r5}; $k_{\rm max}
=\sqrt{4TE}$ is the maximum momentum transfer\cite{r3}, $E$ is the energy of 
the quark; $\alpha $($\beta$)
denotes the included angle between ${\bf v}_p$ and ${\bf k}$ ( ${\bf k_1}$). 
When ${\bf k}_1$ is selected as a polar axis, the approximate analytic 
solutions for Eqs(14) and (15) can be obtained. Because $\left ({{\rm d}W\over
{\rm d}x}\right )^n_{k>k_0}$ is much smaller than $\left ({{\rm d}W\over
{\rm d}x}\right )^n_{k<k_0}$ by numerical estimations, we only give
\begin{equation}
\left ({{\rm d}W\over {\rm d}x}\right )^n_{k<k_0}
          \approx {NC_Q\alpha^2_sm_DI\over 48\pi^2v}
            \left [{3(\pi-2)\over v^2}\ln{\sqrt{4TE}\over m_D}
                  +{(3\pi-2)\over 2}\right ]
\end{equation}
Eq(16) shows $\left ({{\rm d}W\over {\rm d}x}\right )^n$ is positive, 
namely, the parton is accelerated by fluctuations, it 
implies the nonlinear effects are contrary to linear polarization effects,
 The formula still shows
the  energy enhancement of the parton is determined by the fluctuation 
intensity. Considering a case in free fluctuations ($I=4\pi T$)\cite{r8, r9},
for $T=0.25$GeV(critical temperature of deconfinement transition), 
$E=10$GeV, we have
\begin{equation}
\vline {\left ({{\rm d}W\over {\rm d}x}\right )^n\over
\left ({{\rm d}W\over {\rm d}x}\right )^l}\vline\approx 10^{-1}
\end{equation}
Although the acceleration effect is 1 order of magnitude smaller than 
the loss effect
in free fluctuations, QGP produced by heavy ion collisions, in fact, 
is a turbulent system. Since the turbulent fluctuations are much stronger than
free fluctuations, the acceleration effect of parton is quite evident.

\indent
In summary, we calculate the energy variation of a parton due to the nonlinear
polarization effect of QGP and find the parton can be accelerated by fluctuations.

\end{document}